\documentclass[aps,prl,twocolumn,groupedaddress]{revtex4}

\usepackage{graphicx}
\usepackage{amsmath}
\usepackage{float}

\begin{document}


\title{Timing Recollision in Nonsequential Double Ionization by Intense Elliptically Polarized Laser Pulses}
 

\email[]{Kang@atom.uni-frankfurt.de}

\author{H. Kang,$^{1,2*}$ K. Henrichs,$^{1}$ M. Kunitski,$^{1}$ Y. Wang,$^{2}$ X. Hao,$^{3}$ K. Fehre,$^{1}$ A. Czasch,$^{1}$ S. Eckart,$^{1}$ L.Ph.H. Schmidt,$^{1}$ M. Sch\"{o}ffler,$^{1}$ T. Jahnke,$^{1}$ X. Liu,$^{2}$ and R. D\"{o}rner$^{1}$}
\affiliation{$^{1}$Institut f$\ddot{u}$r Kernphysik, Goethe
Universit$\ddot{a}$t Frankfurt, 60438 Frankfurt am Main, Germany
\\$^{2}$ State Key Laboratory of Magnetic
Resonance and Atomic and Molecular Physics, Wuhan Institute of
Physics and Mathematics, Chinese Academy of Sciences, Wuhan 430071,
China\\$^{3}$Institute of Theoretical Physics and Department of
Physics,State Key Laboratory of Quantum Optics and Quantum Optics
Devices, Collaborative Innovation Center of Extreme Optics, Shanxi
University, Taiyuan 030006, China}




\begin{abstract}
We examine correlated electron and doubly charged ion momentum spectra from strong field double ionization of Neon employing intense elliptically polarized laser pulses. An ellipticity-dependent asymmetry of correlated electron and ion momentum distributions has been observed. Using a 3D semiclassical model, we demonstrate that our observations reflect the sub-cycle dynamics of the recollision process. Our work reveals a general physical picture for recollision-impact double ionization with elliptical polarization, and demonstrates the possibility of ultrafast control of the recollision dynamics.
\end{abstract}


\maketitle

The recollision scenario, which is the keystone of strong field physics, describes the process that an electron firstly tunnels out of a Coulomb potential, which is distorted by a strong laser field, and then is accelerated and driven back by the laser field to recollide with its parent ion \cite{CorkumPRL1993}. This process is responsible for many characteristic strong field phenomena such as high-order harmonic generation (HHG), high-order above-threshold ionization (HATI) and nonsequential double ionization (NSDI). Among them, NSDI is of particular interest and has continued to receive intense experimental and theoretical attention (for reviews, see \cite{DonerAdvat2002, BeckerRMP2012}) because it is regarded as one dramatic manifestation of electron-electron correlation in nature. The recollision phenomenon has been discovered by observing a strong enhancement in the double ionization yield occurring for certain laser intensity ranges \cite{HuillierPRA1983}. This enhancement - a characteristic ``knee"-structure, which contradicted the sequential tunneling model - has been observed in all rare gas atoms \cite{FittinghoffPRL1992, WalkerPRL1994, AugstPRA1995} and some molecules \cite{GuoPRA1998, AlnaserPRL2003, CornaggiaPRA2000, ZeidlerPRL2005, EreminaPRL2004}. The probability for recollision is maximal for linear polarized light and decreases strongly with ellipticity. Consequently the ratio of double to single ionization is known to drop with ellipticity \cite{DietrichPRA1994}. 

In this paper we study double ionization as function of the ellipticity of the driving field and show that this allows to answer in more detail at which time the recollision-induced ionization occurs. This information is hidden to experiments with linearly polarized light. In addition, the conceptual simple elliptical light form allows manipulating the recollision in a simple and transparent way. More complex tailored laser fields have already been used to control recollision successfully (see, e.g., \cite{ZhangPRL2014, ChaloupkaPRL2016, EckartPRL2016, MancusoPRL2016}).

\begin{figure}[b]
\includegraphics[width=3.2in]{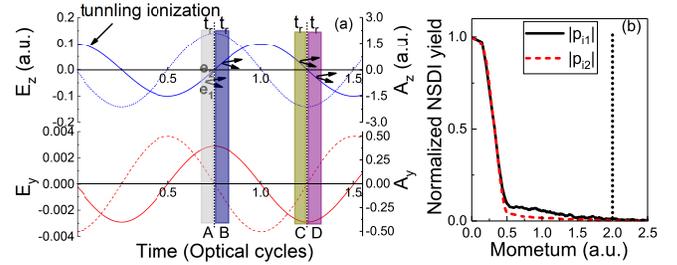}
\caption{\label{Fig1}
(Color online) (a) Time information of the correlated electron emission with elliptically polarized light at 788 nm with a peak intensity of $5\times10^{14}$ W/cm$^{2}$. The electric field (solid curves) and vector potential (dashed curves) along the major axis (the $z$-axis) and the minor axis (the $y$-axis) are plotted as a function of time. The ellipticity is 0.25 here. Note the different scales of the electric fields and vector potentials. The boxes A, C and B, D show that the recollision occurs at $t_{r}$ before and after the $E_{z}$ field zero-crossing (vertical dotted lines), respectively. (b) Calculated probability distribution of the post-recollision momentum of each electron $\left | p_{i1}\right |$ and $\left | p_{i2}\right |$. The dotted vertical line denotes the value of peak vector potential ($\sim$2 a.u., corresponding roughly to the vector potential $\left |A(t_{r})\right |$ at the recollision time $t_{r}$) for the peak intensity of $5\times10^{14}$ W/cm$^{2}$. Other laser parameters are the same as panel (a). See text for details. }
\end{figure}

To gain the maximum information on the dynamics of double ionization with elliptical light we have performed fully differential measurements. A sketch of our experimental strategy is shown in Fig. 1(a). Considering double ionization by an elliptically polarized electric field $\textbf{E}(t)=(0,-\frac{E_{0}}{\sqrt{1+\epsilon^{2}}}\epsilon\sin\omega t,\frac{E_{0}}{\sqrt{1+\epsilon^{2}}}\cos\omega t)$ with the amplitude $E_{0}$, laser frequency $\omega$, and a small ellipticity $\epsilon$, the first electron tunnels out along the major axis (i.e., the $z$-axis) slightly after $t=0$ so that it can recollide with the parent ionic core. A classical analysis \cite{WuPRA2013} has demonstrated that recollision occurs around $nT+3T/4$ ($n=0,1,2...$ and $T$ denotes the optical cycle) or $mT+T/4$ ($m=1,2,3...$ ). For simplicity, we only show recollision time $t_{r}$ within 1.5$T$ in Fig. 1(a). Upon recollision, the second electron may be ionized directly (recollision impact ionization) or be promoted to an excited bound state and then be freed by the laser field (recollision excitation with subsequent ionization) at a later time \cite{FeuersteinPRL2001}. We choose Neon as a target, since for Neon double ionization proceeds mainly via recollision impact ionization \cite{EreminaJPB2003} and the more complicated case of recollision excitation plays a minor role. For recollision impact double ionization, as illustrated in Fig. 1(a), double ionization occurs shortly after the electron-electron collision so each electron's drift velocity, which is due to the acceleration by the field, is determined mostly by the vector potential at the recollision time. In addition the two electrons carry the post-recollision momenta $p_{i1}$ and $p_{i2}$ which arise from the dynamic energy sharing mediated by the electron-electron interaction during the recollision \cite{FeuersteinPRL2001}. The final momenta of the correlated electrons are thus $\textbf{p}_{1}\approx \textbf{p}_{i1} -\textbf{A}(t_{r})$ and $\textbf{p}_{2}\approx \textbf{p}_{i2} -\textbf{A}(t_{r})$ (atomic units are used throughout this paper). Here $\textbf{p}_{i1}$ and $\textbf{p}_{i2}$ satisfy the condition $E_{exc}=({p_{i1}}^{2}+{p_{i2}}^{2})/2$ where $E_{exc}$ is the energy difference between the recolliding electron and the ionization potential of the singly charged ion. Because the recollision time is around the $E_{z}$ field zero-crossing, the value of $A(t_{r})$ is close to the peak vector potential. For the laser peak intensity of $5\times10^{14}$ W/cm$^{2}$, $\left | p_{i1}\right |$ and $\left | p_{i2}\right |$ are much smaller than $\left |A(t_{r})\right |$, which is verified by our semiclassical calculation shown in Fig. 1(b). This establishes a connection between the final electrons' momenta and the recollision time allowing one to experimentally access the sub-cycle dynamics of the recollision process. If the recollision occurs around the $E_{z}$ zero-crossing $nT+3T/4$ or $mT+T/4$ [cases A+B or C+D in Fig. 1(a)], both the $z$-components of the final momenta of the two electrons will have negative or positive values, i.e., $p_{1z}<0, p_{2z}<0$ or $p_{1z}>0, p_{2z}>0$, respectively. While both the $y$-components of the final momenta of the two electrons will shift to negative and positive (or positive and negative) values, if the recollision occurs before and after the $E_{z}$ zero-crossing $nT+3T/4$ (or $mT+T/4$), corresponding to cases A and B (or C and D) in Fig. 1(a), respectively. By momentum conservation ion momentum mirrors the sum momentum of the both electrons and is therefore a powerful observable to unveil the details of the recollision (see, e.g., \cite{MoshammerPRL2000, WeberPRL2000}).

To realize the strategy described above, one has to account for the influence of parent ion's Coulomb potential on the correlated electron emission. This is fully incorporated in a versatile 3D semiclassical model \cite{YePRL2008, WangPRA2017}, which we use in this work together with experimental data for various ellipticities, to demonstrate that the temporal properties of recollision is indeed closely related to the correlated electron momentum distributions also when the Coulomb potential comes into play.

In our experiment, a commercial Ti:Sapphire femtosecond laser system (100 kHz, 100 $\mu$J, 45 fs, Wyvern-500, KMLabs) was employed to generate intense laser pulses at a central wavelength of 788 nm. We used a quarter-wave plate to produce the elliptically polarized pulses. The laser beam was focused by a spherical concave mirror ($f=60$ mm) onto a cold supersonic Ne gas jet. The laser peak intensity in the interaction region was determined by measuring the ``donut''-shape momentum distribution of singly charged Ne$^{+}$ ions with circular polarization \cite{AlnaserPRA2004}. The uncertainty of the peak intensity is estimated to be $\pm20$\%.

\begin{figure}
\includegraphics[width=3in]{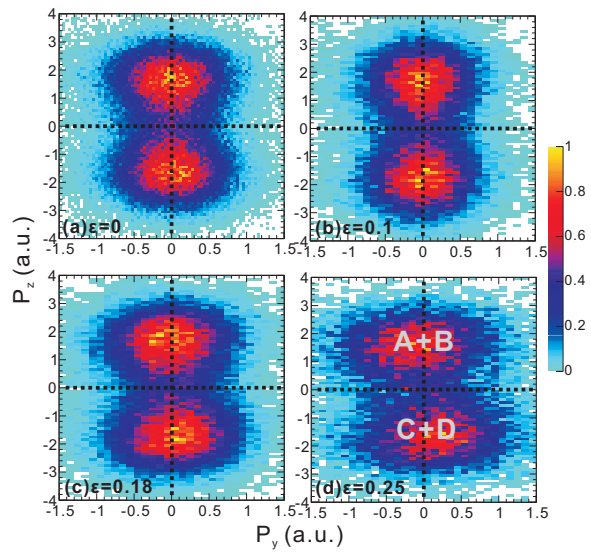}
\caption{\label{Fig2}
(Color online)  Measured momentum distributions of Ne$^{2+}$ ions in the $y$-$z$ polarization plane of elliptically polarized laser pulses at a peak intensity of $5\times10^{14}$ W/cm$^{2}$ and a central wavelength of 788 nm. The ellipticities are varied from 0 to 0.25. The major and minor axes are along the $z$ and $y$ axes, respectively. The color scales have been normalized for comparison purposes.}
\end{figure}

A Cold Target Recoil Ion Momentum Spectroscopy (COLTRIMS) reaction microscope \cite{UllrichRPP2003} has been used to measure the three-dimensional momentum distributions of the doubly charged Ne ion and one of the emitted electrons in coincidence. The details of this setup can be found elsewhere \cite{KevinPRL2013}. We used a half-wave plate to make sure that the major axis of the elliptically polarized light was oriented along the time-of-flight direction (i.e., along the symmetry axis of the spectrometer). To avoid dead-time problems of the particle detectors, the measurement was restricted to the momenta of one of the electrons emitted and the doubly charged ion. The momentum of the other electron was deduced via momentum conservation.

\begin{figure}
\includegraphics[width=2.5in]{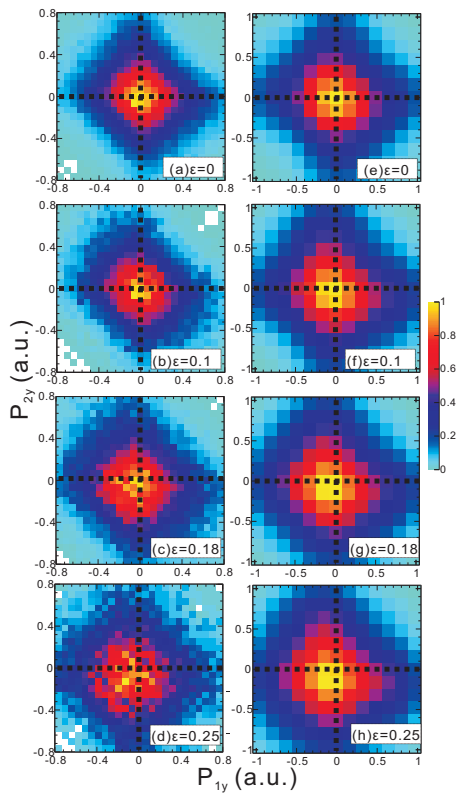}
\caption{\label{Fig3}
(Color online)  Experimental (a)$\sim$(d) and calculated (e)$\sim$(h) correlated electron momentum distributions along the $y$-axis (i.e., the minor axis of the elliptical polarization) for various ellipticities. The laser parameters are the same as Fig. 2. The momenta of both electrons along the $z$-axis are restricted to positive values. The electron pairs shown in this figure correspond to the third and fourth quadrants of Fig. 2. The color scales of the panels have been normalized for comparison purposes.}
\end{figure}

In Fig. 2 we display the measured momentum distributions of doubly charged Ne ions with elliptical polarization at a peak intensity of $5\times10^{14}$ W/cm$^{2}$ for the ellipticities from 0 to 0.25. Over this ellipticity range the ratio R of Ne$^{2+}$/Ne$^{1+}$ drops drastically ($R=6.3\times10^{-4}, 4.0\times10^{-4}, 0.9\times10^{-4}, 0.5\times10^{-4}$ for $\epsilon=0, 0.1, 0.18, 0.25$ in our experiments). For each ellipticity, a symmetric distribution of the Ne$^{2+}$ ions in the $z$-direction can be seen, suggesting that the double ionization is dominated by recollision impact ionization under our experimental conditions (see \cite{BeckerRMP2012}). Cases A+B and C+D shown in Fig. 1(a) correspond to the Ne$^{2+}$ ions located in the 1st+2nd and 3rd+4th quadrants, respectively, as labeled in Fig. 2(d). Examining panels (a) to (d) reveals an increase of accumulation of Ne$^{2+}$ ions in the 2nd and 4th quadrants with increasing ellipticities. The asymmetry between the 1st and 2nd (or the 3rd and 4th quadrants) is due to different probabilities of the recollision occurring before and after the $E_{z}$ field zero-crossing $nT+3T/4$ (or $mT+T/4$). In the following we will select electron pairs in the 3rd and 4th quadrants (C+D) to clarify this point. 

In Figs. 3(a)$\sim$3(d) we present the measured correlated electron momentum distributions along the $y$-axis with the condition that $p_{1z}>0, p_{2z}>0$ for various ellipticities, corresponding to the third and fourth quadrants of Fig. 2 (C+D). From Fig. 3 we can see that with increasing ellipticity, more and more electron pairs become located in the third quadrant. The asymmetry of the electron pairs in the first and third quadrants is in accordance with the asymmetry of the Ne$^{2+}$ ions in the third and fourth quadrants of Fig. 2. 

To simulate our data, we have performed a 3D semiclassical model calculation. This semiclassical model has been successfully used to explain various strong-field double ionization phenomena, e.g., the important role of Coulomb potential \cite{YePRL2008, WangPRA2017}, and its computational details can be found elsewhere \cite{HaoPRA2009}. The calculated results are shown in Figs. 3(e)$\sim$3(h). The observed ellipticity-dependent asymmetry of the electron pairs in the first and third quadrants is well reproduced by the calculation. The discrepancy in the momentum values possibly arises from the fact that the actual peak intensity in the laser focus ($\pm20$\% uncertainty of the peak intensity calibration) could be lower than the one used in the calculation.

To gain insight into the dynamics causing the asymmetry pattern in Fig. 3, we have performed a back analysis approach in our simulations, which allows us to evaluate the probability distributions of the recollision time for specified electron trajectories \cite{time}. Note that the Coulomb potential effects on the electron trajectories are intrinsically included in our simulation. Here we compare the trajectories that contribute to the electron pairs in the first quadrant and the third quadrant of Fig. 3. For comparison purposes, we also present the results for the sum of these two types of trajectories. The calculated results are displayed in Fig. 4. In the calculation, we employ the elliptically polarized electric field $\textbf{E}(t)=(0,-\frac{E_{0}}{\sqrt{1+\epsilon^{2}}}f(t)\epsilon\sin\omega t,\frac{E_{0}}{\sqrt{1+\epsilon^{2}}}f(t)\cos\omega t)$ where $f(t)$ is the pulse envelop which is a constant equal to 1 for the first 10 cycles and exponentially reduced to 0 with 3-cycle ramp. Therefore, only electron recollisions occurring within ($mT$, $mT+T/2$) lead to electron pairs with $p_{1z}>0,p_{2z}>0$.

\begin{figure}
\includegraphics[width=3.2in]{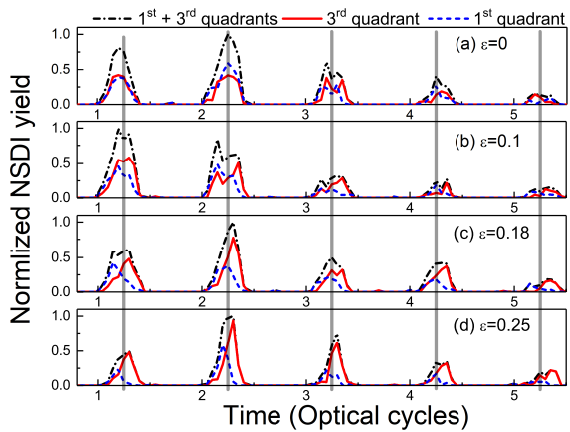}
\caption{\label{Fig4}
(Color online)  Probability distributions of recollision time for various ellipticities. The black dash-dotted curves denote the calculated results for the electron trajectories leading to the electron pairs located in the first and third quadrants of Fig. 3. The red solid (blue dashed) curves denote the results for the electron pairs in the third (first) quadrant. The $E_{z}$ zero-crossing $mT+T/4$ is marked by vertical gray lines. The results have been normalized to the maximum of the black dash-dotted curve for each ellipticity.}
\end{figure}

In the context of the semiclassical model, the contributions to recollision-induced double ionization can be conveniently separated into single-return-collision (SRC) and multiple-return-collision (MRC) trajectories, depending upon whether the recollision occurs when the tunnel-ionized electron returns to the ion for the first time or after passing the ion at least once \cite{HoPRL2005}. For linear polarization $\epsilon=0$, the NSDI probability is expected to decrease rapidly with the travel time of the first electron due to the electronic wave packet's transverse spread. Thus the SRC trajectories should make the dominant contribution. However, the calculated result [black dash-dotted curve in Fig. 4(a)] shows that the second and third peaks become comparable to or even stronger than the first peak. This indicates a significant Coulomb focusing effect in driving these MRC trajectories back to the ionic core. For linear light there is by definition no difference between the first quadrant and the third quadrant [Fig. 3(a)] thus the red and blue curves in Fig. 4(a) coincide.

This changes already for small ellipticity $\epsilon=0.1$. Then, the transverse field component will steer the tunnel-ionized electron away from the ionic core in the $y$-direction. In order to return to the core, the electron needs to have a proper transverse velocity right after tunneling ionization, i.e., post-tunneling transverse velocity. This effect largely suppresses the interaction between the electron and the ionic Coulomb potential. Since the contribution of the MRC electrons to NSDI strongly depends on the Coulomb focusing effect, their probabilities will drop faster so that the SRC electrons dominate the contribution to the NSDI [black dash-dotted curves in Fig. 4(b)]. With further increased ellipticity, however, the contribution of MRC electrons becomes dominant [black dash-dotted curves in Figs. 4(c) and 4(d)]. This is because for higher ellipticities, the corresponding post-tunneling transverse velocities of the SRC electrons need to be significantly larger than that of the MRC electrons, leading to the suppressed contribution of the SRC trajectories \cite{ShilovskiPRA2008}. This effect has been observed in experiments on HATI spectra \cite{SalieresScience2001, LaiPRL2013}. More importantly, the red solid and the blue dashed curves in Figs. 4(c)$\sim$4(d) reveal that, with increasing ellipticity, the recollisions occurring after the $E_{z}$ zero-crossing become more and more important in contributing to the electron pairs in the third quadrant. Consequently, the observed asymmetry pattern in Fig. 3 indicates that the recollisions are more likely to occur after the $E_{z}$ zero-crossing for higher ellipticities. Therefore, the detailed analysis on recollision time distributions supports that we experimentally accessed the recollision process on a sub-femtosecond timescale and implies possibilities of ultrafast control by varying the ellipticity.

\begin{figure}
\includegraphics[width=3.2in]{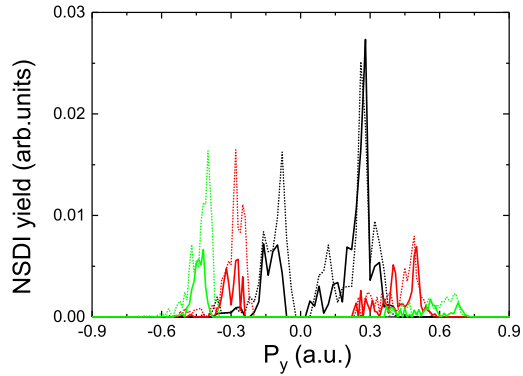}
\caption{\label{Fig5}
(Color online) Probability distributions of the post-tunneling transverse momentum (along the $y$-axis) of the tunnel-ionized electron for ellipticities $\epsilon=0.1$ (black), 0.18 (red), and 0.25 (green), respectively. The solid and dotted lines represent the electron trajectories for the recollision before and after the $E_{z}$ zero-crossing $mT+T/4$, respectively. The results for 0.18 and 0.25 have been multiplied by a factor of 2.7 and 6.0 for visual convenience, respectively.}
\end{figure}

The importance of the post-tunneling momentum for the recollision is further highlighted in Fig. 5, which shows the calculated transverse momentum (along the $y$-axis) distributions of the electron right after tunneling ionization. Here we have analysed the same electron trajectories as used in Fig. 4. We compare the trajectories with recollision occurring before and after the $E_{z}$ zero-crossing $mT+T/4$. The peaks with negative and positive post-tunneling $p_{y}$ correspond to the electrons tunneled from the first half and the second half of the laser cycle, respectively. The calculation shows that for small ellipticity $\epsilon=0.1$, the post-tunneling $p_{y}$ is very small. For higher ellipticities $\epsilon=0.18$ and 0.25, the post-tunneling $p_{y}$ has larger values, as discussed above. Furthermore, for the recollisions occurring after the $E_{z}$ zero-crossing for each ellipticity, the electrons need to have smaller post-tunneling $p_{y}$. According to the tunneling theory \cite{DeloneJOSAB1991}, the probability of the tunnel-ionized electron $\omega(t_{0},\upsilon_{\perp})\sim \upsilon_{\perp} \exp[-2(2I_{p1})^{3/2}/ 3\left| E(t_{0}) \right|] \exp[-\upsilon_{\perp}^{2}(2I_{p1})^{1/2}/ \left| E(t_{0}) \right|]$ ($I_{p1}$ denotes the first ionization potential of Ne) decreases exponentially with the increase of post-tunneling transverse velocity $\upsilon_{\perp}$ \cite{NoteDK}. The larger the ellipticity, the more suppressed is the contribution of the electron trajectories for the recollision before the $E_{z}$ zero-crossing (Fig. 5). Thus more recollisions will occur after the $E_{z}$ zero-crossing with increasing ellipticity.

In summary, we experimentally studied the correlated electron and doubly charged ion momenta from strong field double ionization of Ne by elliptically polarized light. An ellipticity-dependent asymmetry of the correlated electron pair and ion momenta has been observed. With the help of a 3D semiclassical model, we find that the correlated electron momentum distributions along the minor axis of elliptical polarization provide access to the sub-cycle dynamics of recollision and distinguish recollisions before and after the field zero-crossing, which presents a novel approach to obtain information about recollision time \cite{ShafirNature2012}. Furthermore, our data demonstrates that the recollision can be steered by varying the ellipticity. This work reveals a general physical picture of double ionization by recollision with elliptical light, and provides insight into possibilities of ultrafast control of recollision.

This work is supported by Deutsche Forschungsgemeinschaft. H. Kang acknowledges support by the Alexander von Humboldt Foundation. X. Liu acknowledges the support from NNSF of China (Grant No. 11334009).


\begin{thebibliography}{99}

\bibitem{CorkumPRL1993} P.B. Corkum, Phys. Rev. Lett. \textbf{71}, 1994 (1993).
\bibitem{DonerAdvat2002} R. D\"{o}rner, Th. Weber, M. Weckenbrock, A. Staudte, M.
Hattass, H. Schmidt-B\"{o}cking, R. Moshammer, and J. Ullrich, Adv. At. Mol. Opt. Phys. \textbf{48}, 1 (2002).
\bibitem{BeckerRMP2012} W. Becker, X. Liu, P. Ho, and J.H. Eberly, Rev. Mod. Phys. \textbf{84}, 1011 (2012).
\bibitem{HuillierPRA1983} A. l'Huillier, L.A. Lompre, G. Mainfray, and C. Manus, Phys. Rev. A  \textbf{27}, 2503 (1983).
\bibitem{FittinghoffPRL1992} D.N. Fittinghoff, P.R. Bolton, B. Chang, and K.C. Kulander, Phys. Rev. Lett. \textbf{69}, 2642 (1992).
\bibitem{WalkerPRL1994} B. Walker, B. Sheehy, L.F. DiMauro, P. Agostini, K.J. Schafer, and K.C. Kulander, Phys. Rev. Lett. \textbf{73}, 1227 (1994).
\bibitem{AugstPRA1995} S. Augst, A. Talebpour, S.L. Chin, Y. Beaudoin, and M. Chaker, Phys. Rev. A \textbf{52}, R917 (1995); A. Talebpour, C-Y. Chien, Y. Liang, S. Larochelle, and S.L. Chin, J. Phys. B \textbf{30}, 1721 (1997); S. Larochelle, A. Talebpour, and S.L. Chin, J. Phys. B \textbf{31}, 1201 (1998).
\bibitem{GuoPRA1998} C. Guo, M. Li, J.P. Nibarger, and G.N. Gibson, Phys. Rev. A \textbf{58}, R4271 (1998); C. Guo and G.N. Gibson, Phys. Rev. A \textbf{63}, 040701 (2001).
\bibitem{AlnaserPRL2003} A.S. Alnaser, T. Osipov, E.P. Benis, A. Wech, B. Shan, C.L. Cocke, X.M. Tong, and C.D. Lin, Phys. Rev. Lett. \textbf{91}, 163002 (2003).
\bibitem{CornaggiaPRA2000} C. Cornaggia and Ph. Hering, Phys. Rev. A \textbf{62}, 023403 (2000).
\bibitem{ZeidlerPRL2005} D. Zeidler, A. Staudte, A.B. Bardon, D. M. Villeneuve, R. D\"{o}rner, and P.B. Corkum, Phys. Rev. Lett. \textbf{95}, 203003 (2005).
\bibitem{EreminaPRL2004} E. Eremina, X. Liu, H. Rottke, W. Sandner, M.G. Sch\"{a}tzel, A. Dreischuh, G.G. Paulus, H. Walther, R. Moshammer, and J. Ullrich,  Phys. Rev. Lett. \textbf{92}, 173001 (2004).
\bibitem{DietrichPRA1994} P. Dietrich, N. H. Burnett, M. Ivanov, and P. B. Corkum, Phys. Rev. A \textbf{50}, R3585 (1994).
\bibitem{ZhangPRL2014} L. Zhang, X. Xie, S. Roither, Y. Zhou, P. Lu, D. Kartashov, M. Sch\"{o}ffler, D. Shafir, P.B. Corkum, A. Baltu\u{s}ka, A. Staudte, and M. Kitzler, Phys. Rev. Lett. \textbf{112}, 193002 (2014).
\bibitem{ChaloupkaPRL2016} J.L. Chaloupka and D.D. Hickstein, Phys. Rev. Lett. \textbf{116}, 143005 (2016).
\bibitem{EckartPRL2016} S. Eckart, M. Richter, M. Kunitski, A. Hartung, J. Rist, K. Henrichs, N. Schlott, H. Kang, T. Bauer, H. Sann, L.Ph.H. Schmidt, M. Sch\"{o}ffler, T. Jahnke, and R. D\"{o}rner, Phys. Rev. Lett. \textbf{117}, 133202 (2016).
\bibitem{MancusoPRL2016} C.A. Mancuso, K.M. Dorney, D.D. Hickstein, J.L. Chaloupka, J.L. Ellis, F.J. Dollar, R. Knut, P. Grychtol, D. Zusin, C. Gentry, M. Gopalakrishnan, H.C. Kapteyn, and M. M. Murnane, Phys. Rev. Lett. \textbf{117}, 133201 (2016).
\bibitem{WuPRA2013} M.Y. Wu, Y.L. Wang, X.J. Liu, W.D. Li, X.L. Hao, and J. Chen, Phys. Rev. A \textbf{87}, 013431 (2013).
\bibitem{FeuersteinPRL2001}  B. Feuerstein, R. Moshammer, D. Fischer, A. Dorn, C.D. Schr\"{o}ter, J. Deipenwisch, J.R. Crespo Lopez-Urrutia, C. H\"{o}hr, P. Neumayer, J. Ullrich, H. Rottke, C. Trump, M. Wittmann, G. Korn, and W. Sandner, Phys. Rev. Lett. \textbf{87}, 043003 (2001).
\bibitem{EreminaJPB2003} E. Eremina, X. Liu, H. Rottke,W. Sandner, A. Dreischuh, F. Lindner, F. Grasbon, G. G. Paulus, H. Walther, R. Moshammer, B. Feuerstein, and J. Ullrich, J. Phys. B \textbf{36}, 3269 (2003).
\bibitem{MoshammerPRL2000}  R. Moshammer, B. Feuerstein, W. Schmitt, A. Dorn, C. D. Schroter, J. Ullrich, H. Rottke, C. Trump, M. Wittmann, G. Korn, K. Hoffmann, and W. Sandner, Phys. Rev. Lett. \textbf{84}, 447 (2000).
\bibitem{WeberPRL2000} T. Weber, M. Weckenbrock, A. Staudte, L. Spielberger, O. Jagutzki, V. Mergel, F. Afaneh, G. Urbasch, M. Vollmer, H. Giessen, and R. Dorner, Phys. Rev. Lett. \textbf{84}, 443 (2000).
\bibitem{YePRL2008} D.F. Ye, X. Liu, and J. Liu, Phys. Rev. Lett. \textbf{101}, 233003 (2008).
\bibitem{WangPRA2017} Y.L. Wang, S.P. Xu, Y.J. Chen, H.P. Kang, X.Y. Lai, W. Quan, X.J. Liu, X.L. Hao, W.D. Li, S.L. Hu, J. Chen, W. Becker, W. Chu, J.P. Yao, B. Zeng, Y. Cheng, and Z.Z. Xu, Phys. Rev. A \textbf{95}, 063415 (2017).
\bibitem{AlnaserPRA2004} A.S. Alnaser, X.M. Tong, T. Osipov, S. Voss, C.M. Maharjan, B. Shan, Z. Chang, and C.L. Cocke, Phys. Rev. A \textbf{70}, 023413 (2004).
\bibitem{UllrichRPP2003} J. Ullrich, R. Moshammer, A. Dorn, R. D\"{o}rner, L.P.H. Schmidt, and H. Schmidt-B\"{o}cking, Rep. Prog. Phys. \textbf{66}, 1463 (2003).
\bibitem{KevinPRL2013} K. Henrichs, M. Waitz, F. Trinter, H. Kim, A. Menssen, H. Gassert, H. Sann, T. Jahnke, J. Wu, M. Pitzer, M. Richter, M. S. Sch\"{o}ffler, M. Kunitski, and R. D\"{o}rner, Phys. Rev. Lett. \textbf{111}, 113003 (2013).
\bibitem{HaoPRA2009} X.L. Hao, G.Q. Wang, X.Y. Jia, W.D. Li, J. Liu, and J. Chen, Phys. Rev. A \textbf{80}, 023408 (2009).
\bibitem{time} The recollision time here is defined as the time of closest approach of the two electrons after tunneling.
\bibitem{HoPRL2005} P.J. Ho, R. Panfili, S.L. Haan, and J.H. Eberly, Phys. Rev. Lett. \textbf{94}, 093002 (2005).
\bibitem{ShilovskiPRA2008} N.I. Shvetsov-Shilovski, S.P. Goreslavski, S.V. Popruzhenko, and W. Becker, Phys. Rev. A \textbf{77}, 063405 (2008).
\bibitem{SalieresScience2001} P. Sali\`{e}res \emph{et al.}, Science \textbf{292}, 902 (2001).
\bibitem{LaiPRL2013} X.Y. Lai, C.L. Wang, Y.J. Chen, Z.L. Hu, W. Quan, X.J. Liu, J. Chen, Y. Cheng, Z.Z. Xu, and W. Becker, Phys. Rev. Lett. \textbf{110}, 043002 (2013).
\bibitem{DeloneJOSAB1991} N.B. Delone and V.P. Krainov, J. Opt. Soc. Am. B. \textbf{8}, 1207 (1991).
\bibitem{NoteDK} This tunneling formula gives the probability of the first electron (tunnel-ionized electron) trajectories in the semiclassical model. The second electron's (bound electron) initial condition is determined by assuming it in the ground state of Ne$^{+}$. Double ionization occurs due to the recollision of the first electron with the ionic core. By selecting the recolliding electron trajectories, we ruled out the ionization of the second electron before recollision in our calculation. Therefore the double ionization probability is related to the tunneling probability of the first electron. 
\bibitem{ShafirNature2012} Recollision time can also be extracted from high harmonic signal. See, e.g., D. Shafir, H. Soifer, B.D. Bruner, M. Dagan, Y. Mairesse, S. Patchkovskii, M.Yu. Ivanov, O. Smirnova, and N. Dudovich, Nature (London)  \textbf{485}, 343 (2012).

\end{thebibliography}
\end{document}